# George Gamow and Albert Einstein: Did Einstein say the cosmological constant was the "biggest blunder" he ever made in his life?

Galina Weinstein[1]

October 3, 2013

**Abstract:** In 1956/1970 Gamow wrote that much later, when he was discussing cosmological problems with Einstein, he remarked that the introduction of the cosmological term was the "biggest blunder" he ever made in his life. But the cosmological constant rears its ugly head again and again and again. Apparently, Einstein himself has never used the aperçu "biggest blunder"; nevertheless a vast literature grew up around this notion and associated it with Einstein. The present work is prompted by questions put by Mario Livio in his latest book *Brilliant Blunders* as to the phrase "biggest blunder": Did Einstein actually say, "biggest blunder"? I show that in 1947 Einstein wrote Lemaitre that he found it "very ugly" indeed that the field law of gravitation should be composed of two logically independent terms (one of which was the cosmological term). Earlier, in spring 1922 Einstein wrote Max Born that he committed "a monumental blunder some time ago". In 1965 Born commented on this letter: "Here Einstein admits that the considerations which led him to the positive-ray experiments were wrong: 'a monumental [capital] blunder'". It is likely that when Einstein met Gamow he formulated his views in his native German, and perhaps he told Gamow that suggesting his cosmological constant was a "blunder". I suggest that, Einstein perhaps told Gamow that the cosmological constant was a "capital blunder" or a "monumental blunder", and Gamow could have embellished Einstein's words to become the famous aperçu "biggest blunder".

**George Gamow and "biggest blunder"**

George Gamow reported about discussing with Einstein cosmological problems in his **1970** posthumously published autobiography *My World Line*:[1]

"Thus, Einstein's original gravity equation was correct, and changing it was a mistake. Much later, when I was discussing cosmological problems with Einstein, he remarked that the introduction of the cosmological term was the biggest blunder he ever made in his life. But this 'blunder', rejected by Einstein is still used by cosmologists even today, and the cosmological constant demoted by the Greek letter Λ rears its ugly head again and again and again".

---

[1] Written during my stay at the Program for Science and Technology Education, Ben Gurion University, Israel. The year 2013 is Israel's "Space Year".

Scholars explain that "biggest blunder" did not constitute undertaking of Gamow to quote Einstein, but he stringed two words together to represent his memory of a conversation with Einstein. Gamow indeed said: "when I was discussing cosmological problems with Einstein, *he remarked…*" [my emphasis]. John Earman wrote that biggest blunder is "attributed to Einstein by Gamow. Note that Gamow does not purport to be quoting Einstein directly". Gamow says in a *Scientific American* article entitled "The Evolutionary Universe", published in **September 1956**: "Einstein remarked to me many years ago that the cosmic repulsion idea was the biggest blunder he had made in his entire life". Earman explains: "The account given in Gamow's autobiography *My World Line* […] is similar. The existing evidence is insufficient to decide whether Einstein himself used the word 'blunder' or whether this was Gamow's embellishment".[2]

Mario Livio raises the following question: Did Einstein actually say, "Biggest Blunder?" Livio could find no documentation for Einstein saying that the cosmological term was the biggest blunder he ever made in his life. Instead, he claims, all references eventually lead back only to Gamow, who reported Einstein's use of the phrase in the two sources: *My World Line* and the 1956 *Scientific American* article.[3]

Noteworthy, already in 1999 John Stachel told Harvey and Schucking that "The comment ['biggest blunder'] doesn't appear in Einstein's writings".[4]

Livio scrutinized some letters exchanged between Gamow and Einstein and he realized that they both were not close, because the letters were rather "formal". He asks, in effect, why would Einstein use such "strong language" as biggest blunder concerning the cosmological constant with Gamow, and not with his closest friends? He also asks and answers: "Did he think then that the cosmological constant was his 'biggest blunder'? Unlikely. Yes he was uncomfortable with the concept, saying as early as 1919 that it was 'gravely detrimental to the formal beauty of the theory'".[5]

It is often difficult, owing to the lack of surviving evidence, to disentangle the assumptions upon which a phrase rests. Einstein used, to borrow Livio's phrase, "strong language" with colleagues as well as with close friends. Admittedly, the circumstance that gives strength and prominence to a phrase is that authors and Einstein's biographers quote it everywhere such that, consistently with our image of Einstein the magi, it appears to us as if the phrase is indeed strong.

So far as "formal letters" are concerned, we can readily admit that from Einstein's standpoint it was not just a matter of convenience as to whether Einstein sent a letter in English or in German, which was the language he could better communicate with his friends and colleagues. Of course Einstein's German phraseology better represents Einstein's opinions, and appears to us less formal as compared to the English phraseology, which is probably a translation from German to English. Looked at in this way there is difference in kind between the two options of correspondence, the English and German, to give the latter a preference above the former when discussing Einstein's phraseology.

**When did Einstein and Gamow meet?**

So far as Einstein's meeting with Gamow is concerned the evidence indicate that Einstein and Gamow met during World War II (more than 15 years after Einstein dropped the cosmological constant) and shortly afterwards.

Gamow had told somewhat sketchily the following anecdote in *My World Line*: [6]

"There is very little to say about my consultation work for the armed forces of the United States during World War II. It would have been, of course, natural for me to work on nuclear explosions, but I was not cleared for such work until 1948, after Hiroshima. The reason was presumably my Russian origin and the story I had told freely to my friends of having been a colonel in the field artillery of the Red Army at the age of about twenty.

Thus I was very happy when I was offered a consultantship in the Division of High Explosives in the Bureau of ordnance of the US navy Department.

A more interesting activity during that time was my periodic contact with Albert Einstein, who […] served as a consultant for the High Explosive Division. Accepting this consultantship, Einstein stated that because of his advanced age he would be unable to travel periodically from Princeton to Washington, D.C., and back, and that somebody must come to his home in Princeton, bringing the problems with him. Since I happened to have known Einstein earlier, on non-military grounds, I was selected to carry out this job. Thus on every other Friday I took a morning train to Princeton, carrying a briefcase tightly packed with confidential and secret Navy projects.[…]

After the business part of the visit was over, we had lunch either at Einstein's home or at the cafeteria of the Institute for Advanced Study, which was not far away, and the conversation would turn to the problems of astrophysics and cosmology. In Einstein's study there were always many sheets of paper scattered over his desk and on a nearby table, and I saw that they were covered with tensor formulae which seemed to pertain to the unified-field theory, but Einstein never spoke about that. However, in discussing purely physical and astronomical problems he was very refreshing, and his mind was as sharp as ever".

And there was a flutter of surprise, for Livio discovered a small article published in 1986 by Stephen Brunauer entitled "Einstein and the Navy", the scientist who had recruited both Einstein and Gamow to the Navy. In that article, Brunauer described the entire episode in detail. When explaining Gamow's precise role, Brunauer wrote:

"Gamow, in later years, gave the impression that he was the Navy's liaison man with Einstein, that he visited every two weeks, and the professor 'listened' but made no contribution—all false. The greatest frequency of visits was mine, and that was about every two months". Livio mentions that Brunauer visited Einstein at Princeton on **May 16, 1943**. Clearly, Livio says, Gamow exaggerated his relationship with the famous physicist. [7]

It is not surprising that presumably Gamow and of course other people exaggerated their friendship with Einstein. Einstein's fame led people who knew him to write about their personal acquaintance with him. Einstein had a friend, a Hungarian Jewish physician named János Plesch. Jeremy Bernstein writes of Plesch autobiography: [8] "indicates it was there that he began collecting people both as patients and friends. If we can believe his autobiography, he knew everybody", and above all Albert Einstein. Plesch starts the chapter of his autobiography presenting Einstein with the following description, "Among the many scientific men who are, or have been, my friends there is one who out-tops all the others in stature, and that is Albert Einstein".[9] Yet Einstein told him many valuable typical Einstein anecdotes. How do we know whether Plesch's reports are authentic or not? We cross-reference them with letters, unpublished talks, etc., written by Einstein himself.

**Gamow's and Einstein's sense of humor**

Gamow had a well established reputation as a jokester and had been given to hyperbole. Fred Hoyle wrote that Gamow "would shout from the other end: 'the elements were made in less time than you could cook a dish of duck and roast potatoes".[10] However, Einstein also had a great sense of humor. The young and especially the elderly Einstein played tricks a little with biographers, reporters, and people in telling them stories, because they annoyed him with long personal questions. [11]

Gamow fooled Einstein on his 70[th] birthday. In **July 1949** the *Reviews of Modern Physics* devoted an issue to a celebration of Einstein's 70's birthday. It contained articles about Einstein's achievements by Einstein's friends and colleagues. Gamow contributed the paper, "On Relativistic Cosmology", and in this paper he used his litrary gifts to present a lively and humouerous account of his theory: "Another important group of studies based on the general theory of relativity is presented by the work on relativistic cosmology, which is an attempt to understand the development of various characteristic features of our universe as the result of its expansion from the originally homogeneous state. This includes essentially the theory of the origin of atomic species, which presumably took place during the very early epoch when the material forming the universe was in highly compressed and very hot state, and the theory of the formation of galaxies which must have occurred during the later evolutionary period. The neutron-capture theory of the origin of atomic speciaes recently developed by Alpher, Bethe, Gamow, and Delter suggests that different atomic nuclei were formed by the successive aggregation of neutrons and protons which formed the original hot ylem during the early highly compressed stages in the history of the universe".

And in the footnote the references are: G. Gamow, Phys. Rev. 70, 572 (1946); Alpher, Bethe, and Gamow, Phys. Rev. 73, 803 (1948); R. A. Alpher, Phys. Rev. 74, 1577 (1948); R.A. Alpher and R. C. Herman, Phys. Rev. 74, 1737 (1948).[12]

The 1948 αβγ paper was created by Alpher under the supervision of Gamow and was published on **April 1, 1948** in *Physical Review*. April fool's day was Gamow's favorite publication date. Gamow noticed that if he was to add as a co-author Hans Bethe, then the three names would be Alpher, Bethe, Gamow, like alpha beta gamma, even though Hans Bethe, who agreed for his name to be included, had *nothing* to do with that paper.[13] But in his 1949 paper Gamow gave Einstein a birthday present: he wrote about the αβγδ theory, "Deltor", Robert C. Herman, corresponding to the fourth letter in the Greek alphabet. When referring to Robert Herman, Gamow wrote: "R. C. Herman, who stubbornly refuses to change his name to Delter"…

Einstein owned Gamow's humorous popular science books in his personal library in Princeton. For instance one can find two of Gamow's popular books in Einstein's personal library kept in the Einstein Archives:[14]

Gamow, George, *Mr. Tompkins in Wonderland: or stories of c, G, and h* [the three constants of physics], illustrated by John Hookham, 1940, New York: Macmillan. Gamow, George, *One, two, three… infinity: facts & speculations of science*, illustrated by the author, 1947, New York: Viking Press.

The first lines of *One, Two, Three… Infinity* are the verse:[15]

"There was a young fellow from Trinity,
Who took the square root of infinity.
But the number of digits, Gave him the fidgets;
He dropped Math and took up Divinity."

Einstein also invented verses about funny and amusing situations, and about ridiculous incidents. For instance, on the occasion of his 50[th] birthday (14.3.1929) he wrote a fine and marvelous verse to his friends. [16]

**Gamow and the cosmological constant**

An essential point at issue is Gamow's attitude towards the cosmological constant in the **years 1949-1952**. In his above paper "On relativistic Cosmology", published in **the 1949** special volume of *Reviews of Modern Physics* celebrating Einstein 70's birthday, Gamow shortly reviewed a then known problem in cosmology and thereafter dealt with a possible solution to this matter. According to Hubble's measurements of the rate of uniform expansion we assume that it must have started $1.8 \times 10^9$ years ago. At that epoch the material of the universe must have been in a state of very high density and correspondingly high temperature. However, the study of rocks indicated that the solid crust of the earth must have existed for at least $2.5 \times 10^9$ years, so that the age of the universe is closer to $3 \times 10^9$ years. Since this discrepancy is certainly beyond the limits of errors in astronomical measurements, Gamow emphasized that "it appears that we have here the first serious disagreement between the conclusions of relativistic cosmology and the observed facts. This disagreement, however, can be removed by considering the effect of the so-called *cosmological term* in the general equation of the expanding universe".[17]

**In 1952** Gamow explained why the cosmological constant was dropped:[18]

"When the fact that our Universe is not static but is rapidly expanding was recognized, the introduction of cosmological constant became superfluous. However, as we see later, this constant may still be of some help in cosmology, even though the primary reason for its introduction has vanished".

Concerning the "as we see later" in the above excerpt, Gamow called attention to the all-important question of the disagreement between the conclusions of relativistic cosmology and the geological observed facts. He explained that Georges Lemaître introduced the cosmological constant, because the presence of such force would make the universe expand with ever-increasing velocity and shift the position of the zero point in time. Indeed if the expansion process is accelerated, the recession velocities of the neighboring galaxies would have been smaller in the past than today, so that the date of the beginning would be shifted back in time. Assuming such a small numerical value as $10^{-33}$ sec$^{-1}$ for the cosmological constant, one could bring Hubble's original value into agreement with the geological estimate.[19]

But Einstein held that Lemaître was founding too much upon the field equations with the cosmological constant.[20]

And there was a twist to the tale, because **after 1952** Gamow quietly dropped the cosmological constant and never returned to it.[21] In that case, it seems reasonable to expect that Einstein might have influenced Gamow to drop the cosmological constant. According to a report by Victor Alpher, Ralph Alpher – a physics PhD student of Gamow – and Gamow visited Einstein at various points (independently) during the course of World War II and in the post-war period to discuss their work. Einstein's comments (signed) on Alpher's theoretical developments are housed in the Alpher Papers.[22]

Lastly, remember Gamow wrote **in 1956** that Einstein remarked to him "*many years ago* that the cosmic repulsion idea was the biggest blunder he had made in his entire life" [my italics].[23] It seems to the present writer reasonable that, although Einstein and Gamow met occasionally during World War II and shortly afterwards, Einstein had probably spoken about the cosmological constant with Gamow in **the early**

**1940s**. **After 1952**, Gamow dropped the cosmological constant and subsequently, **in 1956**, he wrote his memory of the meeting with Einstein.

**Einstein's phraseology**

One may never know exactly what Einstein told Gamow. It is quite obvious that Einstein and Gamow met sometime in the early 1940s during the war. We cannot say *a priori* anything further about the contents of these meetings beyond Gamow's reports.

Remember in 1970 Gamow reported that: "the cosmological constant demoted by the Greek letter $\Lambda$ rears its ugly head again and again and again". We notice firstly, **in 1947**, Einstein referred to the modified field equations with the $\Lambda$ term in terms of repulsive, disgusting or ugly. On **September 26, 1947**, Einstein wrote back to Lemaître in response to his letter from **July 30, 1947**: [24]

"Since I have introduced this $\Lambda$ term, I had always a bad conscience. But at that time I could see no other possibility to deal with the fact of the existence of a finite mean density of matter. I found it very ugly indeed that the field law of gravitation should be composed of two logically independent terms, which are connected by addition. About the justification of such feelings concerning logical simplicity it is difficult to argue. I cannot help to feel it strongly and I am unable to believe that such an ugly thing should be realized in nature".

We have no assurance that anything of this sort was ever said when Einstein met Gamow, but perhaps it might be possible to interpret Gamow's above excerpt as describing how Einstein felt towards the cosmological term. After all without the cosmological term $\Lambda$ general relativity involves no general constants ("logical simplicity"?). With $\Lambda$ it acquires a constant that at first sight seems to be different from all other constants of physics. [25]

**In 1922** Einstein wrote Max Born: [26]

"I too committed a monumental blunder some time ago (my experiment on the emission of light with positive rays), but one must not take it too seriously. Death alone can save one from making blunders".

More than 40 years later in 1965, Max Born commented on this letter:[27]

"Here Einstein admits that the considerations which led him to the positive-ray experiments were wrong: 'a monumental blunder' [a "capital blunder" in Born's German original comment]. I should add that now (1965), when I read through the old letters again, I could not understand Einstein's observation at all and found it untenable before I had finished reading".

It is likely that when Einstein met Gamow he formulated his views in his native German, and perhaps he told Gamow that suggesting his cosmological constant was a "blunder". I suggest the view, unreasonable as it may seem to some scholars, and they may not altogether agree with it that, Einstein perhaps told Gamow that the cosmological constant was a "capital blunder"; he might have told Gamow that the $\Lambda$ term was a "monumental blunder", and Gamow could have embellished Einstein's words to become the famous aperçu: "biggest blunder he ever made in his life".

**Notes and References**

[1] Gamow, George, *My World Line*, 1970, Viking Press, p. 44.

[2] Earman, John, "Lambda: The Constant that Refuses to Die", *Archives for History of Exact Sciences* 55, 2001, pp. 189-220; p. 189; Gamow, George, "The Evolutionary Universe", *Scientific American* 195, 1956, pp. 66-67.

[3] Livio, Mario, *Brilliant Blunders: From Darwin to Einstein - Colossal Mistakes by Great Scientists That Changed Our Understanding of Life and the Universe*, New York: Simon & Schuster, 2013, p. 232, pp. 236-237, p. 241.

[4] Harvey, Alex and Schucking, Engelbert, "Einstein's mistake and the cosmological constant", *American Journal of Physics* 68, August, 2000, pp. 723-727; p. 726.

[5] Livio. 2013, p. 232, pp. 236-237, p. 241; Rebecca J. Rosen, "Einstein Likely Never Said One of His Most Oft-Quoted Phrases", *The Atlantic*, August 9, 2013.

[6] Gamow, 1970; Gamow, George, "George Gamow – Memoir", in French, A.P. (ed), *Einstein A Centenary Volume*, 1979, London: Heinemann for the International Commission on Physics Education, pp. 29-30.

[7] Livio, 2013, pp. 235-236; Rosen, 2013.

[8] Bernstein, Jeremy, "Janos Plesch Brief life of an unconventional doctor: 1878-1957", *Harvard magazine*, January-February 2004.

[9] Plesch, János, *János. Ein Arzt erzählt sein Leben*, 1949, Paul List Verlag, München / Leipzig/Plesch, John, *János, The Story of a Doctor*, translated by Edward Fitzgerald, 1949, New York, A.A. WYN, INC, p. 101.

[10] Harvey and Schucking, 2000, p. 726; Hoyle, Fred, "Mathematical theory of the origin of matter", *Astrophysics and Space Science*, 198, 1992, pp. 195-230; p.197.

[11] Already in 1919, Einstein told a *New York Times* correspondent (who came to his Berlin home to interview him) about the 1907 man falling from the roof thought experiment. The correspondent transformed the thought experiment from Bern to Berlin and into a realistic amusing story. Presumably Einstein told the correspondent the story in this way, and the latter did not notice that Einstein was fooling him. Nevertheless, the reader can be assured that the imaginary man fell from an imaginary roof in Bern and not in Berlin: "It was from his lofty library, in which this conversation took place, that he observed years ago a man dropping from a neighboring roof – luckily on a pile of soft rubbish – and escaping almost without injury. This man told Dr. Einstein that in falling he experienced no sensation commonly considered as the effect of gravity, which, according to Newton's theory, would pull him down violently toward the earth". "Einstein Expounds his New Theory", *New York Times* 1919, December 3.

[12] Gamow, George, "On relativistic Cosmology", *Reviews of Modern Physics*, 21, 1949, pp. 367-373; p. 369.

[13] Livio, 2013, pp. 167-168.

[14] Einstein Archives, Hebrew university Jerusalem.

[15] Gamow, George, *One, Two, Three, ... Infinity*, 1947, New York: Viking Press.

[16] "Jeder zeiget sich mir heute/ Von der allerbesten Seite/ Und von nah und fern die Lieben/ Haben rührend mir geschrieben/ Und mit allem mich beschenkt,/ Was sich so ein Schlemmer denkt,/ Was für den bejahrten Mann/ Noch in Frage kommen kann./ Alles naht mit süßen Tönen,/ Um den Tag mir zu verschönen, / Selbst die Schnorrer ohne Zahl Widmen mir ihr Madrigal. / Drum gehoben fühl ich mich / Wie der stolze Adlerich. / Nun der Tag sich naht dem End, / Mach ich euch mein Kompliment; / Alles habt ihr gut gemacht / Und die liebe Sonne lacht". Quoted in Seelig Carl, *Albert Einstein; eine dokumentarische Biographie*, 1954, Zürich: Europa Verlag, pp. 215-216.

[17] Gamow, 1949, p. 376.

[18] Gamow, George, *The Creation of the Universe*, New York: Viking Press, 1952, pp. 25-26.

[19] Gamow, 1952, pp. 29-30.

[20] Einstein, Albert, "Remarks to the Essays Appearing in this Collective Volume", in Schilpp, Paul Arthur (ed.), *Albert Einstein: Philosopher-Scientist*, 1949, La Salle, IL: Open Court, pp. 663-688; pp. 684-685.

[21] Kragh, Helge, "George Gamow and the 'Factual Approach' to Relativistic Cosmology", in Kox, Anne J. and Eisenstaedt, Jean (ed) *The Universe of General Relativity*, Einstein Studies, Vol. 11, Boston: Birkhäuser, 2005, p. 181.

[22] https://ralphalpher.com/ ;Alpher, Victor. S., "Ralph A. Alpher, Robert C. Herman, and the Cosmic Microwave Background Radiation", *Physics in Perspective,* September, 2012, 14, 300-334. Victor Alpher is Ralph's son.

[23] Gamow, 1956, pp. 66-67.

[24] "Ich fand es widerwärtig, annehmen zu müssen, dass die Gleichung für das Gravitationsfeld aus zwei logisch unabhängigen Termen zusammengesetzt sein sollte, die sich zueinander additiv verhalten. Es ist schwierig, Argumente zur Rechtfertigung solcher Gefühle zu geben, wie ich sie bezüglich der logischen Einfachheit empfinde. Ich kann mich aber nicht dagegen wehren, sie mit aller Kraft zu empfinden und ich bin nicht im Stande zu glauben, dass eine so widerwärtige Sache in der Natur verwirklicht werden könnte". Einstein to Lemaitre, 26 September 1947, *Einstein Archives* Doc. 15 085. Kragh, Helge, *Cosmology and Controversy: The Historical Development of Two Theories of the Universe*, Princeton: Princeton University Press, 1996, p. 54.

[25] McCrea, William, Hunter, "The Cosmical Constant", *Quarterly Journal of the Royal Astronomical Society* 12, 1971, pp.140-153; p. 146.

[26] Auch ich habe vor einiger Zeit einen monumentalen Bock geschossen (Experiment über Lichtemission mit Kanalstrahlen). Aber man muß sich trösten. Gegen das Böcke-Schießen hilft nur der Tod". Einstein to Born, undated (sometime in spring 1922), Einstein, Albert and Born, Max, *The Born-Einstein Letters, Correspondence between Albert Einstein and Max and Hedwig Born from 1916 to 1955 with Commentaries by Max Born*, translated by Irene Born, Britain: MacMillan, 1971, Letter 42, p. 71; Einstein, Albert and Born, Max, *Albert Einstein Max Born BriefWechsel 1916-1955*, 1969/1991, Austria/New-York: Nymphenburger, 1969/1991, brief 42, p. 102.

[27] "Hier gesteht Einstein, daß seine Überlegung, die zu dem Kanalstrahlenexperiment führte, falsch sei; ein kapitaler Bock. Dazu muß ich sagen, daß ich jetzt (1965), als ich die alten Briefe wieder las, Einsteins Betrachtung überhaupt nicht verstand und sofort für unhaltbar hielt, ehe ich noch weitergelesen hatte". Einstein and Born, Born's comment (English translation) to letter 42, 1971, p. 71; Einstein and Born, Born's Comment (German) to letter 42, 1969/1991, p. 103.